# The Coupling of a Linearized Gravitational Wave to Electromagnetic Fields and Relevant Noise Issues[*]

Fang-Yu LI, Zhang-Han WU, Yi ZHANG

*Department of Physics, Chongqing University, Chongqing 400044*

*According to electrodynamical equations in curved spacetime we consider the coupling of a linearized weak gravitational wave (GW) to a Gaussian beam passing through a static magnetic field. It is found that unlike the properties of the "left-circular" and "right-circular" waves of the tangential perturbative photon fluxes in the cylindrical polar coordinates, the resultant effect of the tangential and radial perturbations can produce the unique nonvanishing photon flux propagating along the direction of the electric field of the Gaussian beam. This result might provide a larger detecting space for the high-frequency GWs in GHz band. Moreover, we also discuss the relevant noise issues.*

PACS: 04.80Nn,   04.30Nk,   04.30Db

In some previous works [1-3] we considered the electromagnetic (EM) response of a Gaussian beam passing through a static magnetic field to the high-frequency gravitational waves (HFGWs). Expected HFGWs include two major classes of the GWs. Ones are the high-frequency relic GWs expected by quintessential inflationary models [4], other ones are the HFGWs predicted by possible laboratory sources [3, 5]. Since the frequencies of such HFGWs can reach up to $10^9$ Hz or higher, they are just the best resonant frequency bands in smaller EM detectors.

In Refs. [1-3], considering symmetry of the Gaussian beam, we used cylindrical polar coordinates and investigated only the tangential perturbative EM energy fluxes (or in the quantum language: perturbative photon fluxes) produced by the HFGW. These tangential perturbative photon fluxes were expressed as the "left circular waves" and "right circular waves" around the symmetrical axis of the symmetrical axis of the

[*] Supported by the National Nature Science Foundation of China Grant No. 10175096 and Hupei Province Key Laboratory Foundation of Gravitational and Quantum Physics under Grants No. GQ0101.

Gaussian beam, and the tangential perturbative photon fluxes generated by the state of $\times$ polarization of the GW are the unique nonvanishing ones passing through the plane $\phi=\pi/2$ or $\phi=3\pi/2$. Thus any photons measured from such perturbative photon fluxes in the above special planes will be a signal of the EM perturbation produced by the HFGW. In fact a circular polarized GW can not only produce the tangential perturbation, but also generate the radial perturbation, and they have some important differences.

In this letter we will consider the radial perturbation together with the tangential perturbation in Cartesian coordinates, and we found a new and important property and a more clear physical picture: the photon fluxes propagating along the $y-$ and $z-$axises (i.e., the directions of components of the magnetic field of the Gaussian beam) will contain both the background and perturbative ones, while the photon flux propagating on the $x-$axis (i.e., the direction of the electric field of the Gaussian beam) is only pure perturbative photon flux. This means that in addition to the planes $\phi=\pi/2$ and $3\pi/2$ in the cylindrical polar coordinates [1-3] (i.e., $yz-$plane in the Cartesian coordinates), all other planes which parallel with $yz-$plane, will be "receiving" planes of the pure perturbative photon flux propagating along the $x-$axis. In this case the detecting space of the HFGW would be greatly increased.

The Supposing that Cartesian component of the electric fields of the Gaussian beam is [6]

$$E_x = \psi = \frac{\psi_0}{\sqrt{1+(z/f)^2}} \exp(-\frac{r^2}{W^2}) \exp\left\{i\left[(k_e z - \omega_e t) - \arctan\frac{z}{f} + \frac{k_e r^2}{2R} + \delta\right]\right\}, \quad (1)$$

where $r^2 = x^2 + y^2$, $k_e = 2\pi/\lambda_e$, $f = \pi W_0^2/\lambda_e$, $W = W_0\left[1+(z/f)^2\right]^{\frac{1}{2}}$, $R = z + f^2/z$, $\psi_0$ is the amplitude of the electric (or magnetic) field of the Gaussian beam, $W_0$ is the spot radius at the plane $z=0$, $\delta$ is an arbitrary phase factor. Supposing that the electric field of the Gaussian beam is pointed along the direction of $x-$axis, that it is expressed as Eq. (1), and that a static magnetic field pointing along the $y-$axis is localized in the region $-l/2 \leq z \leq l/2$, i.e.,

$$B^{(0)} = \hat{B}^{(0)} = \begin{cases} \hat{B}_y^{(0)} & (-l/2 \leq z \leq l/2), \\ 0 & (z \leq -l/2 \text{ and } z \geq l/2). \end{cases} \quad (2)$$

From Eq. (1) and the scalar Helmholtz equation, we have (we use MKS units)

$$n_z^{(0)} = \frac{\langle S^{z\,(0)}\rangle}{\hbar\omega_e} = \frac{\psi_0^2}{2\mu_0\hbar\omega_e^2[1+(z/f)^2]}\left[k_e + \frac{k_e r^2(f^2-z^2)}{2(f^2+z^2)^2} - \frac{f}{f^2+z^2}\right]\exp(-\frac{2r^2}{W^2}), \quad (3)$$

$$n_r^{(0)} = \frac{\langle \overset{(0)}{S^r} \rangle}{\hbar \omega_e} = \frac{\psi_0^2 k_e r \sin^2 \phi}{2\mu_0 \hbar \omega_e^2 [1+(z/f)^2](z+f^2/z)} \exp(-\frac{2r^2}{W^2}), \tag{4}$$

$$n_\phi^{(0)} = \frac{\langle \overset{(0)}{S^\phi} \rangle}{\hbar \omega_e} = \frac{\psi_0^2 k_e r \sin(2\phi)}{4\mu_0 \hbar \omega_e^2 [1+(z/f)^2](z+f^2/z)} \exp(-\frac{2r^2}{W^2}), \tag{5}$$

where $n_z^{(0)}$, $n_r^{(0)}$ and $n_\phi^{(0)}$ represent the average values of the background axial, radial and tangential photon flux densities, respectively; $\langle \overset{(0)}{S^z} \rangle$, $\langle \overset{(0)}{S^r} \rangle$ and $\langle \overset{(0)}{S^\phi} \rangle$ are corresponding the average values of the background EM energy flux densities [2], superscript 0 denotes the background parameters, $\wedge$ stands for the static field, the angular brackets denote the average over time. Because of the nonvanishing $\langle \overset{(0)}{S^r} \rangle$ and $\langle \overset{(0)}{S^\phi} \rangle$, the Gaussian beam will be asymptotically spread as $|z|$ increases. Transforming $n_z^{(0)}$, $n_r^{(0)}$ and $n_\phi^{(0)}$ into the Cartesian coordinates, we get

$$n_x^{(0)} = n_r^{(0)} \cos\phi - n_\phi^{(0)} \sin\phi = 0, \tag{6}$$

$$n_y^{(0)} = n_r^{(0)} \sin\phi + n_\phi^{(0)} \cos\phi = \frac{\psi_0^2 k_e y}{2\mu_0 \hbar \omega_e^2 [1+(z/f)^2](z+f^2/z)} \exp(-\frac{2r^2}{W^2}). \tag{7}$$

According to the electrodynamical equations in curved spacetime, both the Gaussian beam and the static magnetic field will be perturbed by the GW. As shown in Refs. [1, 2, 7], the amplitude ratio of the two first-order perturbative EM fields is approximately $h\tilde{B}^{(0)} / h\hat{B}^{(0)}$, where $\tilde{B}^{(0)}$ is the magnetic field of the Gaussian beam, the notation ~ stands for the time-dependent EM fields. In our case, we have chosen $\tilde{B}^{(0)} \sim 10^{-3}$ T, $\hat{B}^{(0)} = 10$ T, i.e., roughly their ratio is only $10^{-4}$. Thus the former can be neglected. In other words, the contribution of the Gaussian beam is only expressed as the coherent synchroresonance (i.e., $\omega_e = \omega_g$) of it with the first-order perturbative EM fields generated by the direct interaction of the GW with the static field $\hat{B}_y^{(0)}$. Assuming that a circular polarized GW propagates along the $z-$axis, i.e.,

$$\begin{aligned} h_{xx} &= -h_{yy} = h_\oplus = A_\oplus \exp[i(kz-\omega_g t)], \\ h_{xy} &= h_{yx} = h_\otimes = iA_\otimes \exp[i(kz-\omega_g t)]. \end{aligned} \tag{8}$$

In this case, using the first-order perturbative EM energy flux densities $\langle \overset{(1)}{S^r} \rangle_{\omega_e=\omega_g}$, $\langle \overset{(1)}{S^\phi} \rangle_{\omega_e=\omega_g}$ and $\langle \overset{(1)}{S^z} \rangle_{\omega_e=\omega_g}$ obtained in Ref [2], and setting $\delta = \pi/2$ in Eq. (1) (this is always possible), the first-order perturbative photon flux densities in the Cartesian coordinates can be given by the following values.

(a) Region I ($z \leq -l/2$)

$$n_x^{(1)} = n_y^{(1)} = n_z^{(1)} = 0. \tag{9}$$

(b) Region II ($-l/2 \leq z \leq l/2$)

$$n_x^{(1)} = n_r^{(1)} \cos f - n_f^{(1)} \sin f$$

$$= -\frac{1}{\hbar w_e} \left\{ \frac{A_\otimes \hat{B}_y^{(0)} y_0 k_g y(z+l/2)}{4 m_0 \left[1+(z/f)^2\right]^{1/2} (z+f^2/z)} \sin\left(\frac{k_g r^2}{2R} - \arctan\frac{z}{f}\right) \right.$$

$$+ \frac{A_\otimes \hat{B}_y^{(0)} y_0 y(z+l/2)}{2 m_0 W_0^2 \left[1+(z/f)^2\right]^{3/2}} \cos\left(\frac{k_g r^2}{2R} - \arctan\frac{z}{f}\right)$$

$$+ \frac{A_\otimes \hat{B}_y^{(0)} y_0 y}{4 m_0 \left[1+(z/f)^2\right]^{1/2} (z+f^2/z)} \sin(k_g z) \sin\left(k_g z + \frac{k_g r^2}{2R} - \arctan\frac{z}{f}\right)$$

$$\left. + \frac{A_\otimes \hat{B}_y^{(0)} y_0 y}{2 m_0 k_g W_0^2 \left[1+(z/f)^2\right]^{3/2}} \sin(k_g z) \cos\left(k_g z + \frac{k_g r^2}{2R} - \arctan\frac{z}{f}\right) \right\} \exp(-\frac{r^2}{W^2}), \tag{10}$$

$$n_y^{(1)} = n_r^{(1)} \sin f + n_f^{(1)} \cos f$$

$$= \frac{1}{\hbar w_e} \left\{ \frac{A_\oplus \hat{B}_y^{(0)} y_0 k_g y(z+l/2)}{4 m_0 \left[1+(z/f)^2\right]^{1/2} (z+f^2/z)} \cos\left(\arctan\frac{z}{f} - \frac{k_g r^2}{2R}\right) \right.$$

$$+ \frac{A_\oplus \hat{B}_y^{(0)} y_0 y(z+l/2)}{2 m_0 W_0^2 \left[1+(z/f)^2\right]^{3/2}} \sin\left(\arctan\frac{z}{f} - \frac{k_g r^2}{2R}\right)$$

$$- \frac{A_\oplus \hat{B}_y^{(0)} y_0 y}{4 m_0 \left[1+(z/f)^2\right]^{1/2} (z+f^2/z)} \sin(k_g z) \cos\left(\arctan\frac{z}{f} - \frac{k_g r^2}{2R} - k_g z\right)$$

$$\left. - \frac{A_\oplus \hat{B}_y^{(0)} y_0 y}{2 m_0 k_g W_0^2 \left[1+(z/f)^2\right]^{3/2}} \sin(k_g z) \sin\left(\arctan\frac{z}{f} - \frac{k_g r^2}{2R} - k_g z\right) \right\} \exp(-\frac{r^2}{W^2}), \tag{11}$$

$$n_z^{(1)} = \frac{1}{\hbar w_e} \left\{ \frac{A_\oplus \hat{B}_y^{(0)} y_0 (z+l/2)}{4 m_0 \left[1+(z/f)^2\right]^{1/2}} \left[k_g + \frac{k_g r^2 (f^2 - z^2)}{2(f^2+z^2)^2} - \frac{f}{f^2+z^2}\right] \cos\left(\arctan\frac{z}{f} - \frac{k_g r^2}{2R}\right) \right.$$

$$+ \frac{A_\oplus \hat{B}_y^{(0)} y_0 z(z+l/2)}{4 m_0 f^2 \left[1+(z/f)^2\right]^{3/2}} \left[1 - \frac{2r^2}{W_0^2[1+(z/f)^2]}\right] \sin\left(\arctan\frac{z}{f} - \frac{k_g r^2}{2R}\right)$$

$$- \frac{A_\oplus \hat{B}_y^{(0)} y_0}{4 m_0 k_g \left[1+(z/f)^2\right]^{1/2}} \left[k_g + \frac{k_g r^2 (f^2 - z^2)}{2(f^2+z^2)^2} - \frac{f}{f^2+z^2}\right] \sin(k_g z) \cos\left(k_g z - \arctan\frac{z}{f} + \frac{k_g r^2}{2R}\right)$$

$$+\frac{A_\oplus \hat{B}_y^{(0)} y_0 z}{4m_0 k_g f^2 \left[1+(z/f)^2\right]^{3/2}}\left[1-\frac{2r^2}{W_0^2[1+(z/f)^2]}\right]\sin(k_g z)\sin\left(k_g z - \arctan\frac{z}{f} + \frac{k_g r^2}{2R}\right)\right\}\exp(-\frac{r^2}{W^2}), \quad (12)$$

(c) Region III ($l/2 \le z \le l_0$), where $l_0$ is the size of the effective region in which the second-order EM perturbation retains a plane wave form [2].

$$n_x^{(1)} = -\frac{1}{\hbar w_e}\left\{\frac{A_\otimes \hat{B}_y^{(0)} y_0 k_g ly}{4m_0 \left[1+(z/f)^2\right]^{1/2}(z+f^2/z)}\sin\left(\frac{k_g r^2}{2R} - \arctan\frac{z}{f}\right)\right.$$
$$\left.+\frac{A_\otimes \hat{B}_y^{(0)} y_0 ly}{2m_0 W_0^2 \left[1+(z/f)^2\right]^{3/2}}\cos\left(\frac{k_g r^2}{2R} - \arctan\frac{z}{f}\right)\right\}\exp(-\frac{r^2}{W^2}), \quad (13)$$

$$n_y^{(1)} = \frac{1}{\hbar w_e}\left\{\frac{A_\oplus \hat{B}_y^{(0)} y_0 k_g ly}{4m_0 \left[1+(z/f)^2\right]^{1/2}(z+f^2/z)}\cos\left(\arctan\frac{z}{f} - \frac{k_g r^2}{2R}\right)\right.$$
$$\left.+\frac{A_\oplus \hat{B}_y^{(0)} y_0 ly}{2m_0 W_0^2 \left[1+(z/f)^2\right]^{3/2}}\sin\left(\arctan\frac{z}{f} - \frac{k_g r^2}{2R}\right)\right\}\exp(-\frac{r^2}{W^2}), \quad (14)$$

$$n_z^{(1)} = \frac{1}{\hbar w_e}\left\{\frac{A_\oplus \hat{B}_y^{(0)} y_0 l}{4m_0 \left[1+(z/f)^2\right]^{1/2}}\left[k_g + \frac{k_g r^2 (f^2-z^2)}{2(f^2+z^2)^2} - \frac{f}{f^2+z^2}\right]\cos\left(\arctan\frac{z}{f} - \frac{k_g r^2}{2R}\right)\right.$$
$$\left.+\frac{A_\oplus \hat{B}_y^{(0)} y_0 lz}{4m_0 f^2 \left[1+(z/f)^2\right]^{3/2}}\left[1-\frac{2r^2}{W_0^2[1+(z/f)^2]}\right]\sin\left(\arctan\frac{z}{f} - \frac{k_g r^2}{2R}\right)\right\}\exp(-\frac{r^2}{W^2}). \quad (15)$$

According to Eqs. (3), (6), (7) and (10)-(15), we have

$$n_x = n_x^{(1)}, \quad (16)$$

$$n_y = n_y^{(0)} + n_y^{(1)}, \quad (17)$$

$$n_z = n_z^{(0)} + n_z^{(1)}. \quad (18)$$

Eqs. (3), (6), (7), (10)-(15) and (16)-(18) show that under the synchroresonance condition (i.e., $w_e = w_g$), $n_y$ and $n_z$ contain both the background and first-order perturbative photon fluxes. Thus $n_y^{(1)}$ and $n_z^{(1)}$ will be swamped by $n_y^{(0)}$ and $n_z^{(0)}$, respectively. In other words, $n_y^{(1)}$ and $n_z^{(1)}$ have no observable effect. Unlike $n_y$ and $n_z$, $n_x$ contains only the pure first-order perturbative photon flux $n_x^{(1)}$ (i.e., $n_x^{(0)} = 0$).

Therefore any photon measured from $n_x$ may be a signal of the EM perturbation produced by the GW. It is interesting to compare the perturbative photon flux with ones discussed in Refs.[1, 2, 3]. We can find that in Refs.[1, 2, 3] only the planes $\phi = \pi/2$ and $\phi = 3\pi/2$ in the cylindrical polar coordinates (i.e., the $yz$-plane in the Cartesian coordinates) are the "receiving" surfaces of the purely perturbative photon flux. Unlike the previous works, in addition to the $yz$-plane, any the surfaces which parallel with the $yz$-plane will be "receiving" planes of the pure perturbative photon flux. In this case requirement of the displaying condition to the HFGW can be greatly relaxed.

In Table I we list the pure perturbative photon fluxes passing through some typical "receiving" surfaces, where the typical laboratory parameters are chosen. Namely, $P = 10^5$ W, $10^3$ W and $10$ W respectively (the power of the Gaussian beam); $W_0 = 0.1$ m (the spot radius of the Gaussian beam); $\hat{B}_y^{(0)} = 10$ T (the background static magnetic field); $-l/2 \leq z \leq l_0$ ($l = 0.1$ m, $l_0 = 0.3$ m), i.e. the area of the "receiving" surface is roughly $10^{-2}$ m$^2$. Moreover, $A_\oplus = A_\otimes = 10^{-30}$, $\nu_g = 3$ GHz (they are not only typical orders of the high-frequency relic GWs expected by the quintessential inflationary models [4], but also typical orders predicted by possible laboratory HFGWs sources [3, 5]).

TABLE I  The perturbative photon fluxes passing through some typical "receiving" surfaces

| Positions of The "receiving" surfaces | I. $x = 0$, $0 \leq y \leq W_0$ $-l/2 \leq z \leq l_0$ | II. $x = -W_0$, $0 \leq y \leq W_0$ $-l/2 \leq z \leq l_0$ | III. $x = -2W_0$, $0 \leq y \leq W_0$ $-l/2 \leq z \leq l_0$ | IV. $x = 0$, $W_0 \leq y \leq 2W_0$ $-l/2 \leq z \leq l_0$ |
|---|---|---|---|---|
| $P$ (W) | $n_x^{(1)}$ (s$^{-1}$) | $n_x^{(1)}$ (s$^{-1}$) | $n_x^{(1)}$ (s$^{-1}$) | $n_x^{(1)}$ (s$^{-1}$) |
| $10^5$ | $3.62 \times 10^3$ | $1.98 \times 10^3$ | $1.61 \times 10^2$ | $3.40 \times 10^3$ |
| $10^3$ | $3.62 \times 10^2$ | $1.98 \times 10^2$ | $1.61 \times 10$ | $3.40 \times 10^2$ |
| $10$ | $3.62 \times 10$ | $1.98 \times 10$ | $1.61$ | $3.40 \times 10$ |

Figure 1 gives four typical positions I, II, III, and IV of the "receiving" surfaces. Expect for the position I, the all other "receiving" surfaces II, III, and IV are laid to regions outside the spot radius $W_0$ of the Gaussian beam. Thus the above results have a more realistic meaning to distinguish and display the perturbative photon fluxes.

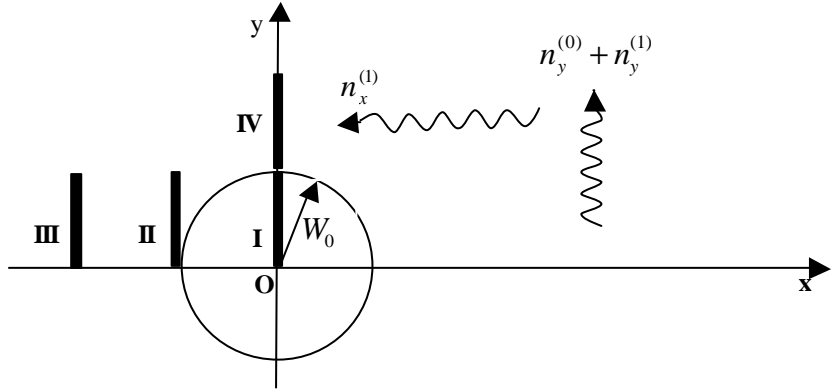

*FIG. 1. The four typical "receiving" surfaces of the perturbative photon fluxes, where I is just laid at yz-plane ($0 \leq y \leq W_0$), while II, III, and IV are all laid to the regions outside the spot radius $W_0$ of the Gaussian beam.*

Although the issues of the signal-to-noise ratio in EM detection systems have been extensively discussed and reviewed, our EM detection system has following obvious advantages to noise problems: (1) Since the resonance frequencies (GHz band) are much higher than ones of usual environment noise (e. g., mechanical, seismic and other ones), the requirements of suppressing such noise can be further relaxed; (2) For the possible external EM noise sources, using a Faraday cage would be useful. Once the EM system is isolated from the outside world by the Faraday cage, possible noise sources would be the remained thermal photons and self-background action; (3) Because of the "random motion" of the remained thermal photons and the highly directional propagated property of the perturbative photon fluxes, the influence of such background would be effectively suppressed in the local regions; (4) Since the background photon fluxes passing through $yz$-plane (including other planes which parallel with the $yz$-plane) vanish, while the perturbative photon flux $n_x^{(1)}$ is the unique nonvanishing photon flux passing through such planes and it has maxima at the $yz$-plane, the signal-to-noise ratio would be much larger than unity at such surfaces, although the perturbation are much less than the background in other regions; (5) Even if for the mixed regions of the signal and the background action, because the self-background action would be decay as $\exp(-\frac{2r^2}{W^2})$ [see, Eqs. (3)-(5)], while the signal would be decay as $\exp(-\frac{r^2}{W^2})$ [see, Eqs. (10)-(15)], the signal-to-noise ratio may reach up to unity in regions of $r = 4W$, roughly. Of course, in such regions the signal will be reduced to a very small value, but increasing background static field (if possible) may be a better way to remove this difficulty, because in this way the number of the background real photons does not change. Moreover, low-temperature and vacuum operation might effectively reduce the frequencies of the remaining thermal photons and avoid dielectric dissipation.

In short the signal and noise in our EM system have a very different physical behavior (e. g., the propagating direction, polarization, distribution and phase) in the local regions, thus the requirement of distinguish the signal and noise may be greatly relaxed. Moreover, if there is a small deviation between the "receiving" surface and $yz$-plane, then using special fractal materials [8], it is possible to reflect the signal and background in different directions and regions, i.e. they can be made as an "equivalent spectroscope" for the signal and background. We will discuss these questions elsewhere.